\documentclass[aps,prd,showpacs,twocolumn]{revtex4}
\usepackage{amssymb}
\usepackage{amsmath}

\begin{document}

\title{Generalized Einstein-Maxwell field equations in the Palatini formalism}

\author{Gin\'{e}s R.P\'{e}rez Teruel$^1$}

\affiliation{$^1$Departamento de F\'{i}sica Te\'{o}rica, Universidad de Valencia, Burjassot-46100, Valencia, Spain}

\date{today}

\begin{abstract}
We derive a new set of field equations within the framework of the Palatini formalism.These equations are a natural generalization of the Einstein-Maxwell equations which arise by adding a function $\mathcal{F}(\mathcal{Q})$, with $\mathcal{Q}\equiv F^{\alpha\beta}F_{\alpha\beta}$ to the Palatini Lagrangian $f(R,Q)$.The result we obtain can be viewed as the coupling of gravity with a nonlinear extension of the electromagnetic field.In addition,a new method is introduced to solve the algebraic equation associated to the Ricci tensor. 
\end{abstract}

\maketitle

\section{Introduction.Palatini $f(R,Q)$theories revisited}
\label{sec:a}

In the last few years,different versions of modified theories of gravity have experimented wide interest in the literature. 
In particular, $f(R)$ Palatini theories have been regarded among the most promising attemps to generalize Einstein's theory of gravity \cite{Ol05},\cite{CaLa},\cite{Ol11},\cite{FeTs},\cite{Lo}.This is because of several reasons. First of all, this formalism provides a very elegant way to derive a cosmological constant and therefore a possibility to explain the observed cosmic speedup.
On the other hand, it exists an intrinsic theoretical interest in the extensions of general relativity (GR),when it comes to incorporate quantum mechanics to a theory of gravity.

However,not only to fully capture the phenomenology of GR, but also for being able to describe a richer variety of physical effects, we should go beyond $f(R)$ theories.We can obtain this within the Palatini approach by just enlarging the class of Lagrangians from $f(R)$ to $f(R,Q)$,where $Q\equiv R_{\mu\nu} R^{\mu\nu}$.\\
The field equations for these class of Lagrangians are, for the metric and the connection respectively \cite{OlAl},\cite{Ol1},\cite{Ol2}

\begin{equation}\label{Ricci}
f_R R_{\mu\nu}-\frac{1}{2}fg_{\mu\nu}+2f_QR_{\mu\alpha}R^{\alpha}_{\nu}=k^2T_{\mu\nu}
\,
\end{equation}

\begin{equation}\label{Connection}
\nabla_\alpha[\sqrt{-g}(f_Rg^{\beta\gamma}+2f_QR^{\beta\gamma})]=0
\,
\end{equation}

with $f_R\equiv \partial_R f $, and $f_Q\equiv\partial_Qf$.

The general algorithm to attack these equations was described in\cite{OlAl}, and consists in several steps. First, we need to find a relation between $R_{\mu\nu}$ and the matter sources. Rewritting (\ref{Ricci}), using $P^{\nu}_{\mu}=R_{\mu\alpha}g^{\alpha\nu}$ we find,

\begin{equation}\label{Ricci_Contracted}
2f_QP^{\alpha}_{\mu}P^{\nu}_{\alpha}+f_RP^{\nu}_{\mu}-\frac{1}{2}f\delta^{\nu}_{\mu}=k^2T^{\nu}_{\mu}
\,
\end{equation}

This can be seen as a matrix equation, which establishes an algebraic relation $P^{\nu}_{\mu}=P^{\nu}_{\mu}(T^{\beta}_{\alpha})$.
Once the solution of (\ref{Ricci_Contracted}) is known, (\ref{Connection})
can be written is terms of $g_{\mu\nu}$ and the matter, which allows to find a solution for the connection by means of algebraic manipulations.The connection can thus be expressed as the Levi-Civita connection of an auxiliary metric $h_{\mu\nu}$ which is related with $g_{\mu\nu}$ by a non-conformal relation \cite{OlAl}

The solution of (\ref{Ricci_Contracted}) is only known in some particular cases, like the perfect fluid or the scalar field, but we point out here for theoretical purposes that a general solution exists and can be found explicitly.
In order to prove this point, let us rewrite (\ref{Ricci_Contracted}) in matrix notation

\begin{equation}\label{matrix_Ricci}
2f_Q\hat{P^2}+f_R\hat{P}-\frac{1}{2}f\hat{I}=k^2\hat{T}
\,
\end{equation}
 
wich has the structure of a quadratic matrix equation, $\hat{A}\hat{X^2}+\hat{B}\hat{X}+\hat{C}=0$.

We can make the following identification,\\

$\hat{A}\equiv 2f_Q\hat{I}$, $\hat{B}\equiv f_R\hat{I}$,  $\hat{C}\equiv -(\frac{f}{2}\hat{I}+k^2\hat{T})$\\

It is straightforward to show that the quadratic matrix equation (4),satisfies all the conditions required \cite{HiKi}.This allow us to complete the square in the usual way to find the following solution for $f_Q\neq 0$

\begin{equation}\label{The Solution}
\hat{P}(\hat{T})=-\frac{f_R}{4f_Q}\hat{I}+\frac{1}{2f_Q}\sqrt{\alpha\hat{I}+\beta\hat{T}}
\,
\end{equation}

where we have defined
\begin{equation}\label{Coefficient_alpha}
\alpha=\frac{1}{4}(f_R^2+4f\cdot f_Q)
\,
\end{equation}

\begin{equation}\label{Coefficient_beta}
\beta=2k^2f_Q
\,
\end{equation}

Therefore, if the matrix $\alpha\hat{I}+\beta\hat{T}$, has a square root, an explicit solution $\hat{P}(\hat{T})$ will always exist, but the mathematical tools required to compute the square root of this matrix will depend on the particular problem chosen.

\newpage

\subsection{Solving for a diagonal matrix}

When $\alpha\hat{I}+\beta\hat{T}$ is diagonal the solution will be automatic.
We are going to see now an example of this type, and in the next subsection we will present a general algorithm to proceed in other, more general, situations.\\The energy momentum tensor of a perfect fluid can be written as

\begin{equation}\label{perfectfluid}
\displaystyle T_{\alpha\beta}=(p+\rho)u_{\alpha}u_{\beta}+pg_{\alpha\beta}
\,
\end{equation}

where $p$ is the pressure of the fluid and $\rho$ its density.Writing the last equation in matrix notation, after a bit of algebra we find the following result

\begin{equation}\label{Diagonal_one}
\displaystyle\alpha\hat{I}+\beta\hat{T} = 
     \begin{pmatrix}
      \alpha-\beta\rho    & \overrightarrow{0} \\
      \overrightarrow{0}  & (\alpha+\beta p)\hat{I}_{3x3}\\
      \end{pmatrix}
\,      
\end{equation}
\\
which is a diagonal matrix, and therefore computing its square root will be straightforward:

\begin{equation}\label{Squareroot}
\displaystyle\sqrt{\alpha\hat{I}+\beta\hat{T}}=
\begin{pmatrix}
      \sqrt{\alpha-\beta\rho}    & \overrightarrow{0}       \\
      \overrightarrow{0}         & (\sqrt{\alpha+\beta p})\hat{I}_{3x3}\\
      \end{pmatrix}
\,      
\end{equation}

In the last expression, it was selected the positive sign of the square roots of the coefficients in order to be consistent with the limit $f_{Q}\rightarrow{0}$. These results allow us to write (5) for a perfect fluid as

\begin{equation}\label{The Solution for Diagonal}
\displaystyle \hat{P}(\hat{T})= 
\begin{pmatrix}
      \Omega               & \overrightarrow{0}  \\
      \overrightarrow{0}   & \omega\hat{I}_{3x3} \\
      \end{pmatrix}
\,      
\end{equation}

where
\begin{equation}\label{Final_Coefficient1}
\Omega=\frac{2\sqrt{\alpha-\beta\rho}-f_R}{4f_Q}
\,
\end{equation}
 
\begin{equation}\label{Final_Coefficient2}
\omega=\frac{2\sqrt{\alpha+\beta p}-f_R}{4f_Q}
\,
\end{equation}

We should point out that all these results are consistent with the formalism developed in \cite{OlAl},\cite{BaOl},\cite{Ol12}for the perfect fluid, but the method described here provides a powerful and direct computation of the matrix $\hat{P}(\hat{T})$, a calculation that in some particular cases may be almost automatic.  

\subsection{A method based in the Schur descomposition for the general case}

Here, we will attempt to provide an algorithm to compute the square root of (5) in the general case.Let us assume that the matrix $\hat{S}\equiv \alpha\hat{I}+\beta\hat{T}$ is nonsingular,which means that its determinant does not vanish. 

If $S \in \mathbf{R}^{4x4}$, then there exists a real orthogonal matrix $\hat{U}$, such that

\begin{equation}\label{Schur1}
\displaystyle\ \hat{U}^{T}\hat{S}\hat{U}=\hat{W}
\,
\end{equation}

where $\hat{W}$ is an upper diagonal matrix, known in linear algebra as the \textbf{Schur form}\cite{Hi87},of the matrix $\hat{S}$.If we can find a matrix $\hat{L}$ which is a square root of $\hat{W}$ ($\hat{L}^2\equiv\hat{W}$), then it can be proved that the matrix

\begin{equation}\label{Schur2}
\displaystyle \hat{J}=\hat{U}\hat{L}\hat{U}^T
\,
\end{equation} 

satisfies the identity $\hat{J}^2\equiv \hat{S}$, and therefore is a square root of $\alpha \hat{I}+\beta \hat{T} $. 
In order to describe the several steps needed to complete the algorithm, we have to make some clarifications.Schur's theorem guarantees that given any square,real matrix $\hat{S}$, the descomposition (14) exists. We first need to compute the eigenvalues and eigenstates of $\hat{S}$, and then we can construct the matrix $\hat{U}$, builded by eigenvectors of $\hat{S}$. The next step requires to compute a square root $\hat{L}$ of the upper diagonal matrix $\hat{W}$. 
To achieve this goal, we can make use of the relation $\hat{L}^2=\hat{W}$, to write for $j\geq i$,

\begin{equation}\label{Schur3}
\sum_{k=1}^{j}L_{ik}L_{kj}=W_{ij}
\, 
\end{equation}
 
This equation can be descomposed into two other equations.First we have

\begin{equation}\label{Schur4}
L_{ii}^{2}=W_{ii}
\,
\end{equation}

and for $j>i$,
\begin{equation}\label{Schur5}
L_{ii}L_{ij}+L_{ij}L_{jj}=W_{ij}-\sum_{k=i+1}^{j-1}L_{ik}L_{jk}
\,
\end{equation} 

Therefore, equation (18) provides an algorithm for computing the remaining blocks $L_{ij}$ once known the diagonal blocks $L_{ii}$.
The final step will be the computation of the square root of $\hat{S}$ by means of the transformation,
$\hat{J}=\hat{U}\hat{L}\hat{U}^T$.

\newpage 
\section{$f(R,Q)$ Lagrangians with a function of the scalar $F^{\mu\nu}F_{\mu\nu}$}
The Palatini action that leads to the field equations (\ref{Ricci}) and (\ref{Connection})
for the metric and the connection is
\begin{equation}\label{action}
S[g,\Gamma,\psi_m]=\frac{1}{2k^2}\int{d^4x\sqrt{-g}f(R,Q)+S[g,\psi_m]}
\,
\end{equation}

where $g_{\alpha\beta}$ is the space-time metric, $\Gamma^{\mu}_{\alpha\beta}$ is the connection, which is indepentent of the metric, $\psi_m$ represents the matter fields, and finally, the scalars $R\equiv g_{\mu\nu}R^{\mu\nu}$, $Q\equiv R_{\alpha\beta}R^{\alpha\beta}$. 
We want to consider here the possibility of replacing in the same action the term $S[g,\psi_m]$ by a function of scalars of the field strength $F^{\mu\nu}$.Taking into account the identity $g_{\mu\nu}F^{\mu\nu}= 0$, due to the fact that $F^{\mu\nu}$ is skewsymmetric, the most natural choice is consider a function $\mathcal{F}(\mathcal{Q})$, with $\mathcal{Q}\equiv F^{\alpha\beta}F_{\alpha\beta}$

With this replacement,we find convenient to write the action as follows:
\begin{equation}\label{action2}
S[g,\Gamma,A]=\frac{1}{2k^2}\int{d^4x\sqrt{-g}[f(R,Q)-k^\prime\mathcal{F}(\mathcal{Q})]}
\,
\end{equation}

where $A_{\mu}$ is a second independent connection, defined by means of $F_{\mu\nu}=\nabla_{\mu}A_{\nu}-\nabla_{\nu}A_{\mu}$, and $k^\prime$ is a constant with the appropiate dimensions.In general, the field strength $F_{\mu\nu}$ in terms of $\Gamma$ and the potential vector $A$ will be

\begin{equation}\label{potentialvector} \nabla_{\mu}A_{\nu}-\nabla_{\nu}A_{\mu}=\partial_{\mu}A_{\nu}-\partial_{\nu}A_{\mu}-(\Gamma^{\rho}_{\mu\nu}-\Gamma^{\rho}_{\nu\mu})A_{\rho}
\,
\end{equation} 

In the following we will assume $\Gamma^{\rho}_{\mu\nu}=\Gamma^{\rho}_{\nu\mu}$ which means that we set the torsion to zero.\\
In these conditions the last term of (21) vanishes and therefore,  $\nabla_{\mu}A_{\nu}-\nabla_{\nu}A_{\mu}\equiv \partial_{\mu}A_{\nu}-\partial_{\nu}A_{\mu}$.\\

We must point out that (20) represents a particular case of Nonlinear Electrodynamics (NED) coupled to gravity.A family of $f(R)$ Palatini Lagrangians coupled to NED have been studied recently in \cite{OlRu}
The variation of the part of the action which contains the term $f(R,Q)$ can be found elsewhere \cite{OlAl},and therefore our task will be restricted to compute the variation of the second part which we will denote as follows

\begin{equation}\label{action3}
\mathcal{S_{\mathcal{Q}}}=\frac{-k^\prime}{2k^2}\int{d^4x\sqrt{-g}\mathcal{F}(\mathcal{Q})}
\,
\end{equation}

This variation gives the following result
\begin{equation}\label{action4}
\delta S_{\mathcal{Q}}=-\frac{k^\prime}{2k^2}\int{d^4x\left[\mathcal{F}(\mathcal{Q})\delta\sqrt{-g}+\sqrt{-g}\frac{\partial{\mathcal{F}}}{\partial{\mathcal{Q}}}\delta\mathcal{Q}\right]}
\,
\end{equation}

Since $\mathcal{Q}=g^{\mu\alpha}g^{\nu\beta}F_{\mu\nu}F_{\alpha\beta}$, it is easy to see that

\begin{equation}\label{action5}
\delta{\mathcal{Q}}=2F_{\mu\alpha}F^{\alpha}_{\nu}\delta g^{\mu\nu}+2F^{\mu\nu}\delta F_{\mu\nu}
\,
\end{equation}

Inserting this last result together with the variation of the determinant of the metric in (23) we get

\begin{eqnarray}
\delta S_{\mathcal{Q}}&=& -\frac{k^\prime}{2k^2}\int d^4x\sqrt{-g}\Big[-\frac{\mathcal{F}}{2}g_{\mu\nu}\delta g^{\mu\nu}\nonumber\\& &+
\mathcal{F}_{\mathcal{Q}}\Big(2F^{\mu\nu}\delta F_{\mu\nu}+2F_{\mu\alpha}F^{\alpha}_{\nu}\delta g^{\mu\nu}\big)\big]
\,
\end{eqnarray}

where $\mathcal{F}_{\mathcal{Q}}\equiv \frac{\partial{\mathcal{F}}}{\partial{\mathcal{Q}}}$. 

The next step requires to express $\delta F_{\mu\nu}$ in terms of $\delta A_{\mu}$ which can be done using the definition of $F_{\mu\nu}$,

\begin{equation}\label{variation_potentialvector}
\delta F_{\mu\nu}=\nabla_{\mu}(\delta A_{\nu})-\nabla_{\nu}(\delta A_{\mu})
\,
\end{equation}

Using the last identity in (25), we obtain after grouping terms:

\begin{eqnarray}
\delta S_\mathcal{Q}&=&-C\int d^4x \sqrt{-g}\Big[\Big(-\frac{\mathcal{F}}{2}g_{\mu\nu} + 2\mathcal{F}_\mathcal{Q}F_{\mu\alpha}F^\alpha_\nu\big)\delta g^{\mu\nu} \nonumber\\& &+ 2\mathcal{F}_\mathcal{Q} F^{\mu\nu}\left(\nabla_{\mu}(\delta A_{\nu})-\nabla_{\nu}(\delta A_{\mu})\right)\big]
\end{eqnarray}

with $\displaystyle C\equiv\frac{k^\prime}{2k^2}$.It is convenient now to split the action into two parts in order to compute properly the last contribution

\small
\begin{equation}
M=-2C\int d^4x\sqrt{-g}\mathcal{F}_{\mathcal{Q}}F^{\mu\nu}\left[\nabla_{\mu}(\delta A_{\nu})-\nabla_{\nu}(\delta A_{\mu})\right]
\end{equation}
\normalsize

Using integration by parts and removing the total derivatives we find 
\begin{eqnarray}
M&=&-2C\int d^4x\Big[-\nabla_{\mu}(\sqrt{-g}\mathcal{F}_{\mathcal{Q}}F^{\mu\nu})\delta A_{\nu}\nonumber\\& &+
\nabla_{\nu}(\sqrt{-g}\mathcal{F}_{\mathcal{Q}}F^{\mu\nu})\delta A_{\mu}\big]
\end{eqnarray}
We have succeeded in expressing the variation of the action $\mathcal{S}_{\mathcal{Q}}$ in terms of variations of $g_{\mu\nu}$ and $A_{\mu}$. This is equivalent to the algorithm used in \cite{OlAl} to express the variation of the action (19) which contains the $f(R,Q)$ Lagrangian, purely in terms of varations of $g_{\mu\nu}$ and the independent connection $\Gamma^{\alpha}_{\mu\nu}$. 

It is possible to write (29) in a more compact form rearranging indices

\begin{equation}\label{action9}
M=-2C\int d^4x \nabla_{\nu}\Big[\sqrt{-g}\mathcal{F}_\mathcal{Q}\left(F^{\mu\nu}\delta^{\lambda}_{\mu}-\delta^{\nu}_{\mu}F^{\mu\lambda}\right)\big]\delta A_{\lambda}
\, 
\end{equation}

With this result, the final expression for $\delta \mathcal{S}_{\mathcal{Q}}$ in (27) will be:

\begin{eqnarray}
\delta S_\mathcal{Q}&=&-C\int d^4x \sqrt{-g}\Big[\Big(-\frac{\mathcal{F}}{2}g_{\mu\nu} + 2\mathcal{F}_\mathcal{Q}F_{\mu\alpha}F^\alpha_\nu\big)\delta g^{\mu\nu} \nonumber\\& &+
\nabla_{\nu}\Big[\sqrt{-g}\mathcal{F}_\mathcal{Q}\left(F^{\mu\nu}\delta^{\lambda}_{\mu}-\delta^{\nu}_{\mu}F^{\mu\lambda}\right)\big]\delta A_{\lambda}\big]
\end{eqnarray}
To obtain our field equations we need to compute the variation of the other contribution to the action in (20).But as we said before, this task has already been done elsewhere \cite{OlAl},\cite{Ol1},and the details will be omitted here.The variation gives:

\begin{eqnarray}
\delta S[g,\Gamma]&=&\frac{1}{2k^2}\int d^4x\Big[\delta g^{\mu\nu}\sqrt{-g}(f_{R}R_{\mu\nu}-\frac{f}{2}g_{\mu\nu}\nonumber\\& &+
2f_{Q}R_{\mu\alpha}R^{\alpha}_{\nu})+\nabla_{\beta}[\sqrt{-g}\Lambda^{\lambda\nu}]\delta\Gamma^{\beta}_{\lambda\nu}\big]
\,
\end{eqnarray}

where $\Lambda^{\mu\nu}\equiv f_{R}g^{\mu\nu}+2f_{Q}R^{\mu\nu}$.
Let us define now the following tensor 
\begin{equation}\label{Generalized_T}
\mathcal{T}_{\mu\nu}\equiv \mathcal{F}_{\mathcal{Q}}F_{\mu\alpha}F^{\alpha}_{\nu}-\frac{\mathcal{F}}{4}g_{\mu\nu}
\,
\end{equation}
Note that when $\mathcal{F}=\mathcal{Q}\equiv F^{\alpha\beta}F_{\alpha\beta}$ it would imply that $\mathcal{F}_{\mathcal{Q}}=1$, and in this case (\ref{Generalized_T}),reduces to the usual definition of the energy-momentum tensor of the electromagnetic field.It is also important to note that in general this tensor is not traceless:

\begin{equation}\label{Trace}
\displaystyle g^{\mu\nu}\mathcal{T}_{\mu\nu}=\mathcal{Q}\mathcal{F}_\mathcal{Q}-\mathcal{F}
\,
\end{equation}

In the context of $f(R)$ theories\cite{OlRu},this fact is directly related with the existence of additional matter terms yielding modified dynamics.
Finally,combining (32) with all our results, we get:

\begin{equation}
f_R R_{\mu\nu}-\frac{1}{2}fg_{\mu\nu}+2f_QR_{\mu\alpha}R^{\alpha}_{\nu}=\frac{k^2}{4\pi}\mathcal{T}_{\mu\nu}
\,
\end{equation}

\begin{equation}
\nabla_\alpha[\sqrt{-g}(f_Rg^{\beta\gamma}+2f_QR^{\beta\gamma})]=0
\,
\end{equation}

\begin{equation}
\nabla_{\beta}[\sqrt{-g}\mathcal{F}_{\mathcal{Q}}F^{\lambda\beta}]=0
\,
\end{equation}
This is a system which reduces to the usual Einstein-Maxwell field equations of GR when $f(R,Q)=R$, and $\mathcal{F}(\mathcal{Q})=\mathcal{Q}$.

\section*{Acknowledgements}

I would like to thank Gonzalo J.Olmo for useful comments and discussions about the Palatini approach.

\end{document}